\documentclass{arxiv}
\usepackage{amssymb}
\usepackage{rotating}
\usepackage{epsfig}
\usepackage{subfigure}
\usepackage{rotating}
\usepackage{adjustbox}
\usepackage{graphicx}
\usepackage{epstopdf}
\usepackage{cite}
\usepackage{caption}
 \usepackage{xcolor}
\usepackage{ulem}

\title{\bf On (Newcomb-)Benford's law: a tale of two papers and of their disproportionate citations.\\
How citation counts can become biased}
\author{\small Tariq Ahmad MIR$^ {1,a,b}$ and Marcel AUSLOOS$^ {2,b,3,c,4,d,5}$\footnote{Deceased}.\\}

\address{\footnotesize
$^1$Nuclear Research Laboratory, Astrophysical Sciences Division, Bhabha Atomic Research Centre, Srinagar-190 006, Jammu and Kashmir, India. \\$^a$ $e$-$mail$ $address$: taarik.mir@gmail.com\\\
$^b$ $e$-$mail$ $address$: tamir@barc.gov.in\\\
$^2$  GRAPES\footnote{Group of Researchers for Applications of Physics in Economy and Sociology} \\ rue de la Belle Jardini\`ere 483, B-4031, Angleur, Li\`ege, Belgium \\ 
$^b$ $e$-$mail$ $address$:  marcel.ausloos@ulg.ac.be\\
$^3$ Department of Statistics and Econometrics, \\The Bucharest University of Economic Studies,  \\ Caeia Dorobantilor 15-17,  010552 Bucharest, Romania \\
$^c$ $e$-$mail$ $address$: marcel.ausloos@csie.ase.ro \\
$^4$ School of Business, University of Leicester,    Brookfield, \\ Leicester, LE2 1RQ, UK    \\
$^d$ $e$-$mail$ $address$: ma683@le.ac.uk \\
 $^5$    Universitatea Babeș-Bolyai, \\ Str. Mihail Kogălniceanu nr. 1, 400084, Cluj-Napoca,  Romania\\
}
\begin{document}

\catchline{}{}{}{}{}

\maketitle


\section{Abstract}
The first digit (FD) phenomenon i.e., the significant digits of numbers in large data are often distributed according to a logarithmically decreasing function was first reported by S. Newcomb and then many decades later independently by F. Benford. 
After its century long neglect the last three decades have seen huge growth in the number of relevant publications. However, notwithstanding the rising popularity the two independent proponents of the phenomenon are not equally acknowledged an indication of which is disproportionate number of citations accumulated by Newcomb (1881) and Benford (1938). 
In the present study we use citation analysis to show that the formalization of the eponym Benford's law, a name questionable itself for overlooking Newcomb's contribution, by Raimi (1976) had a strong  adverse effect on the future citations of Newcomb (1881). 
Furthermore, we identify the papers published over various decades of the developmental history of the FD phenomenon, which latter turned out to be amongst the most cited ones in the field. We find that lack of its consideration, intentional or occasionally out of ignorance for referencing by the prominent papers, is responsible for a far lesser number of citations of Newcomb (1881) in comparison to Benford (1938).

\section{Keywords}

Benford's law; first digit; Simon Newcomb; Frank Benford; citations


\section{Introduction}

American astronomer and mathematician S. Newcomb  
observed that the first few pages of logarithmic table books are somewhat dirtier, are being more thumbed, than are the latter ones. Knowing that logarithmic tables begin with smaller numbers he concluded that users in turn more oftenly look for numbers beginning with smaller digits; he reported his observation as Newcomb (1881) (hereafter referred to as NC), a two page note 
devoid of any data analysis. 

About six decades later physicist and engineer F. Benford, apparently ignorant of NC, reported exactly the same observation, referring to it as the law of anomalous numbers, in Benford (1938) (hereafter referred to as BF), a twenty-two page paper richer in detail. NC was completely forgotten. On the other hand, BF gained early attraction and the observation eventually came to be known, only after Benford, as Benford's law (BL). 

To test his observation Benford had gathered data from twenty fields as diverse as physical constants, atomic and molecular masses, street addresses of American men of sciences, length of rivers etc. and concluded that first significant digits of numbers indeed follow a logarithmic distribution
\begin{equation}
P_{B}(d)= log_{10}(1+\frac{1}{d}), 
\end{equation}
where $P_{B}(d)$ is the probability of a number having  $d$ (= 1, 2, 3..., 9) as the first non-zero digit.

According to equation (1) in a large collection of numerical data the overall distribution of first significant digits (FSD), leftmost non-zero digit, is such that numbers having smaller FSD are far more abundant than those with larger value FSD. For example, the proportion of digit 1 as FSD is 30\% whereas that of digit 9 is only about 4\%. Similarly digit 2 has a proportion of about 17\% whereas digit 8 has only about 5\%. This is against the common intuition according to which each digit from 1 to 9 is expected to appear as the FSD with a uniform proportion of about ($\frac{1}{9}$) 11\% had the appearance of all digits been equally likely.

The theoretical proportions for each digit from 1 to 9 to appear as FSD are as shown in Table 1.
\begin{table}[h]
\tbl{The distribution of first significant digits as predicted by Benford's law}
{\begin{tabular}{@{}ccccccccccc@{}} \toprule
Digit \hphantom{00} & 1 & 2 & 3 & 4 & 5 & 6 & 7 & 8 & 9 \\
Proportion \hphantom{00} & 0.301 & 0.176 & 0.125 & 0.097 & 0.079 & 0.067 & 0.058 & 0.051 & 0.046 \\ \botrule
\end{tabular} \label{ta1}}
\end{table}

Being merely an empirical observation BL hardly gained any attention for more than a century save for an occasional trickle of publications. However,  
during the last three decades, at this time of writing, the number of relevant publications, Benford literature, has grown enormously with 
more than hundred papers being published on a yearly basis. 
The interested reader is referred to the repository ''benfordonline.net`` known as \textbf {Benford Online Bibliography}, hereafter referred to as BOB, a database dedicated to comprehensive coverage of all: news reports, blogs, documentaries, books, doctoral dissertations and research papers, that is happening around the law (Berger et al., 2023).

The entire Benford literature can be broadly categorized as publications (i) verifying its prevalence amongst data from different disciplines (Benford, 1938; Campanario \& Coslado, 2011; Mir, 2012, 2016; Mir et al., 2014; Herteliu et al., 2015; Ausloos et al., 2015, 2017, 2021; Capalbo et al., 2023; Golbeck, 2023), (ii) explaining its mathematical foundations (Pinkham, 1961; Raimi,1976; Hill, 1995a, 1995b, 1995c; Berger \& Hill, 2011; Whyman, 2021), and (iii) advancing its practical applications (Nigrini, 1996b; Riccioni \& Cerqueti, 2018; Cerqueti \& Lupi, 2021).  
Several books (Berger \& Hill, 2015; Kossovsky, 2014; Miller, 2015; Nigrini, 2012) 
highlight important research contributions, made over the years, that have thrown up the FD phenomenon  
to the forefront of research from its initial days of recluse and ridicule. 

NC having been totally forgotten the initial interest in the  FD phenomenon was solely driven by the immediate reception of BF and consequently the latter paper has been an invariable pivot of BL related research. Comparatively the discussions, other than a few brief introductory comments, on NC are only scant despite its appearance being about six decades prior. 
Raimi (1976) the first comprehensive review of the state of research on the FD phenomenon discussed the merits and shortcomings of its many explanations available then. Nigrini (1996a) reviewed decadal contributions made to the theoretical understanding and the application aspects of the law. 

Goudsmith and Furry (1944) gave BF its first ever citation attempting an explanation of the curious observation which in turn saw further expansion by Furry and Hurwitz (1945). Goudsmith (1977) revisited the problem after more than three decades and noted much to his surprise, that about forty papers had already been published on this rather empirical observation that he had taken up earlier only as a welcome diversion from his routine research activity. 
Goudsmith got to know of NC, which by this time had completely been consigned to oblivion, only after Raimi (1976) (Logan \& Goudsmith, 1978) and recalled his own chance encounter with BF when actually a succeeding paper in the journal (Bethe et al., 1938), was of relevance to his work Goudsmith and Saunderson (1940).   
Therefore, Goudsmith (1977) and Logan \& Goudsmith (1978) attribute the quick appreciation of BF 
to its close proximity to Bethe et al. (1938) in the journal. 
 
By analogy Mir and Ausloos (2018) found the foregoing insight convincing since the articles preceding and succeeding NC in the journal have negligibly low citation
count, a testimony of their abysmal scholarly worth. The initial neglect of NC can therefore be attributed 
to its less fortunate position in the journal.

The above discussion begs the interesting question that if the close proximity to an important paper, in some journal, can lead to the appreciation of its seemingly insignificant and unrelated neighbor, how about the effect on the future citations of a given paper, if the subsequently published prominent papers, the most cited ones in the field fail to cite 
the same. NC and BF papers are indeed the fountainhead of everything that is being written about the FD phenomenon and therefore keeping up with the scholarly tradition of fair citation practices one would have expected that contribution of both proponents would be equally acknowledged i.e. cited.  
However, in addition to its initial century long neglect the disdain of NC has continued unabated resulting in accumulation of far lesser number of citations to NC than there are to BF. The assertion is best elucidated by a Google Scholar (GS) search, dated  December 22, 2025, that shows the number of citations to BF is 2776 whereas with only 1483 citations NC is lagging close to halfway behind. Such a large difference between the citations of the two papers is unanticipated in view of the recent explosion of academic interest in the phenomenon. The peculiar citation pattern of disregarding NC can be further appreciated from the fact that Hill (1995a), the third most cited paper with 1090 citations ({\color{blue}{https://scholar.google.com/citations?user=LL0gUwkAAAAJ\&hl=en}}) on GS as on December 22, 2025, is accumulating citations faster than NC and will soon dethrone the latter to become the second most, after BF, cited paper in the field.  

Thus, we observe that the path to popularity, from initial skepticism to a matured academic field of research, for the FD phenomenon,
has been uneven  
and is therefore in itself a subject of interest from the informetric point of view. Study of delayed recognition of the phenomenon has shown that both NC and BF papers are sleeping beauties with the former having been in deep sleep for 110 years and the latter for a relatively lesser period of 31 years. Both were awakened in 1995 by prince Hill (1995a) (Mir \& Ausloos, 2018). The proliferation of Benford literature during the last three decades has been aided by the growth of online databases making data easily accessible for the validation of the law. 
Furthermore, Mir and Ausloos (2018) showed that deep slumber of NC to have been fueled by lack of its acknowledgement in some of the earliest papers, that later on went to be among the most cited ones in the field. But for the lack of consideration, NC could have been woken-up much earlier. 

The more than century long (about one hundred and forty years) neglect of NC is prominently highlighted whenever BL is discussed and in fact has been the starting point for most of the publications on BL. Yet the reasons for such a long neglect of NC have hardly been discussed. In the present study, we bring NC back in sharp focus by digging deeper into the entire volume of Benford literature to understand why the former, notwithstanding its widespread knowledge amongst the concerned researchers, continues to be undermined i.e. is being ignored for citations. 

We show how during various decades of their fascinating citation journey BF was being discussed whereas at the same time NC was getting pushed aside. Study of papers published immediately after the appearance of BF shows that initial disregard of NC was purely out of 
ignorance. Furthermore, it is generally accepted that Newcomb was unjustly overlooked for the naming of the phenomenon as BL. Nevertheless, we examine the bibliography of the publications citing Raimi (1976) and show quantitatively, for the first time, that the formalization of this questionable eponym 
did indeed overshadow importance of NC thereby adversely effecting its future citations.

Further on, we identify papers, some of which are amongst the most cited ones in the field, published immediately after Raimi (1976) that disregarded NC even though being in know of its existence. Moreover, the disdain for NC is found to be more widespread amongst the most cited papers.
We lenghtily investigate their citation histories to show that lack of consideration, citation, of NC inspite of the authors having knowledge of its existence
is responsible for the disproportionate number of citations to NC.



\section{Data}

The data for the present study was obtained from BOB (Berger et al., 2023), an online repository having certain functionalities typical of a regular citation database. Updated regularly BOB is dedicated to track the research on various theoretical and applicational aspects of BL. The articles indexed can be arranged chronologically according to the year of their publication and also alphabetically according to the last names of their respective first authors.  Furthermore, it provides complete reference to the article of interest along with its online information, the articles citing a given article and the articles a given article is citing within its own database. BOB is open-access which is a major advantage over Scopus and Web of Science both of which require paid subscription. On the other hand, cross-referencing of articles within BOB is very advantageous over Google Scholar (GS), the other open-access citation database. Former provides complete bibliography of articles citing any given article whereas the latter lists only the citing articles; to locate their bibliography, access to the full papers is required for which one is again redirected to paid citation services.

\section{Data analysis}

BOB database had 2371 items indexed when accessed on December 22, 2025. In line with our aim of citation analysis we noted down all the papers in the chronological order i.e. according to the date of their publication, in print, as they are listed on BOB. Many of the entries on BOB turned out to be news items, blog posts, weblinks reporting popular commentaries and magazine coverages, articles which normally do not have any bibliography. Further, also listed are some scholarly publications for which no bibliographic information is provided. In the absence of any references, 343 such entries, being of no use in citation analysis, were removed from the data. Thus we are left with 2028 publications with proper information of references; this dataset forms the basis of our study. 

To find out the publications that cite NC and BF we manually inspected the references of each of them by making use of the function ''works that this work references`` on BOB. There are publications that cite neither NC nor BF though they cite other papers relevant to BL; we first noted down the yearly number of such publications. Further, we also noted the yearly number of publications citing at least either of NC and BF or citing both simultaneously. The historical publication record from 1938, the year of BF, up to year 2024 is shown in Fig. 1. Data for the years 2023 and 2024 are expected to be incomplete in view of the time lag between the publication of a paper and its indexing in the database. Data up to year 1980 is shown in inset plot to prevent the congestion of the scale values on the horizontal axis. The general neglect and lack of any interest in BL is readily visible upto year 1980 when only 1-digit number of publications, the highest being seven in the year 1969, 
were being reported on a yearly basis. The year 1981 saw the publication count rise to ten, a 2-digit number, for the first time.

\begin{figure}
\hspace*{-25pt}
\subfigure{\label{}\includegraphics[width=0.7\linewidth, height=1.1\linewidth,  angle=90,]{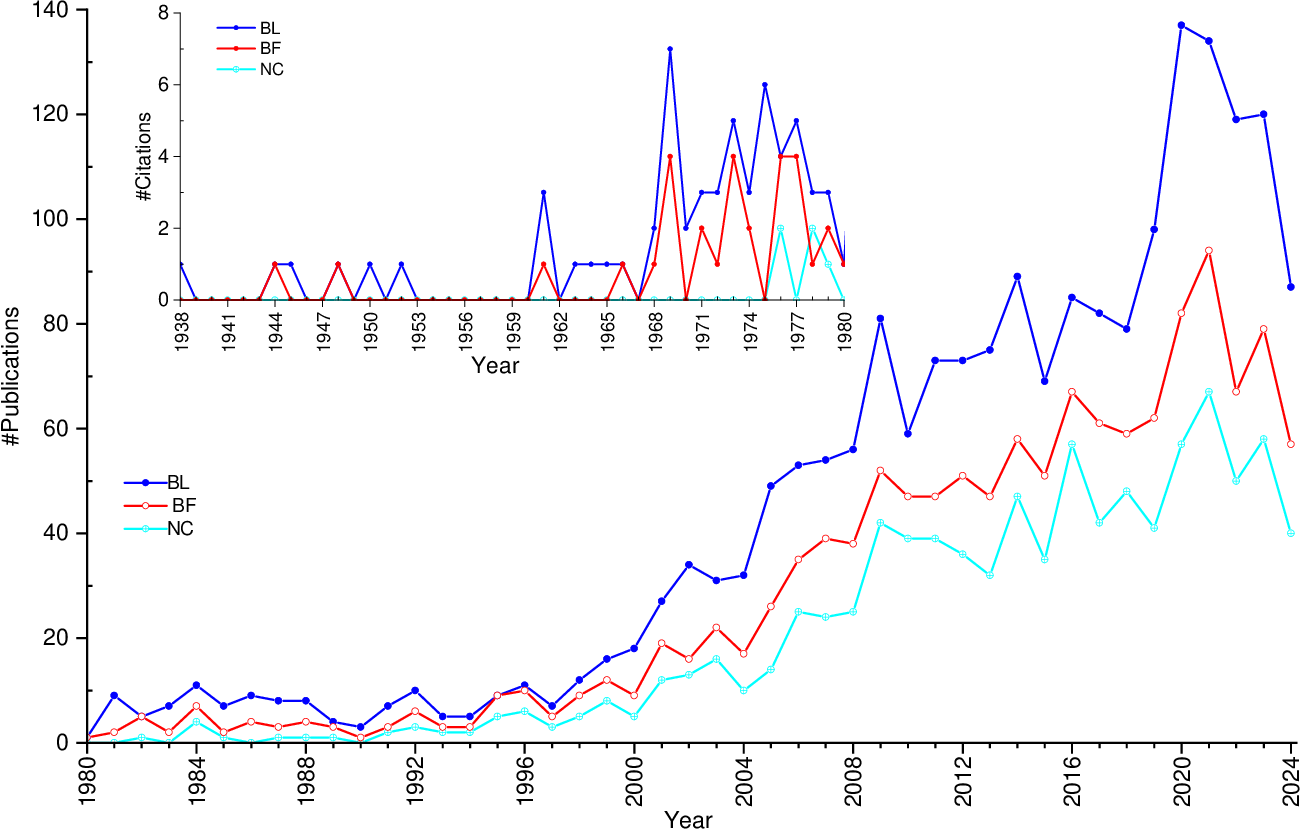}}
\vspace*{-10pt}
\caption{Comparison of yearly publications on BL, those citing Benford (1938) (BF) and Newcomb (1881) (NC). Insert plot is for years 1938-1980.}
\end{figure}

The increase in the yearly number of publications is indicative of rising popularity of BL, but the fact that curve {\color{red}(red color)} corresponding to the publications citing BF is leading over the one {\color{cyan}(cyan color)} citing NC is also indicative of the disproportionate, citations, popularity of the two proponents. For example, 137 papers were published in year 2020 out of which 82 cite BF whereas NC is cited by only 57. Year 2021 saw a total of 134 papers published out of which 94 cited BF whereas NC had only 67 citations. Similarly out of 119 papers that appeared in year 2022 BF is cited by 67 whereas only 50 cite NC. Thus it is clear that more citations are being added to BF than to NC. There are a total of 1325 citations to BF whereas the number for NC is 925. The difference between the two being 400. Thus BF is cited by about 65\% ($\frac{1325}{2028}$)  of the papers whereas as NC is acknowledged, cited, by about 45\% ($\frac{925}{2028}$) of the papers only. In other words about 20\% of the papers do not cite NC at all. 

The disproportionate number of citations to NC is evident from Fig. 1. where the plotted data consists of both the standalone citations and the cocitations of the two papers. Our analysis shows that the latter account for the maximum number of citation counts. 

To further unmask the dismal appreciation of NC we disambiguate the citations of the two papers by finding how many times the two are cited independently of each other. This is elucidated in Fig.2. where we show the yearly number of papers that cite NC but not BF (NC/BF) (ii) BF but not NC (BF/NC) and (iii) those cociting both papers (BF \& NC). It is observed that most often the two are cocited, BF \& NC, followed by BF/NC, standalone citations to BF, and then very far is NC/BF, standalone citations to NC. A casual inspection of the inset plot of Fig. 2. shows that upto 1977 NC is never cited in isolation and  it received, after appearance of BF, its first standalone citation in 1978 (Stigler, 1978) indicated by last tick in the inset plot of Fig. 2. We exemplify the data for same years as explained in case of Fig.1. In year 2020 BF \& NC are cocited 55 times, BF/NC cited 26 times whereas only one publication cited NC/BF. In year 2021 BF \& NC are cocited 65 times, BF/NC received 29 citations whereas NC/BF was cited only four times. Further, in 2022 the two papers were cocited 46 times, BF/NC received 20 citations whereas NC/BF had only 4 citations. Overall the two are cocited 877 times, BF/NC has received 441 citations whereas the same for NC/BF is distant value of 49 citations. Thus BF is cited independently  of NC in about 21\% ($\frac{441}{2085}$) of the published papers whereas NC is rarely cited 2\% ($\frac{49}{2085}$)  independently of BF. Whenever NC is cited BF is invariably cited along. But the reverse is not true that is to say when BF is cited NC is not necessarily to be cited. This in turn indicates that NC is acknowledged, cited, only because of BF.

\begin{figure}
\hspace*{-25pt}
\subfigure{\label{}\includegraphics[width=0.7\linewidth, height=1.1\linewidth,  angle=90,]{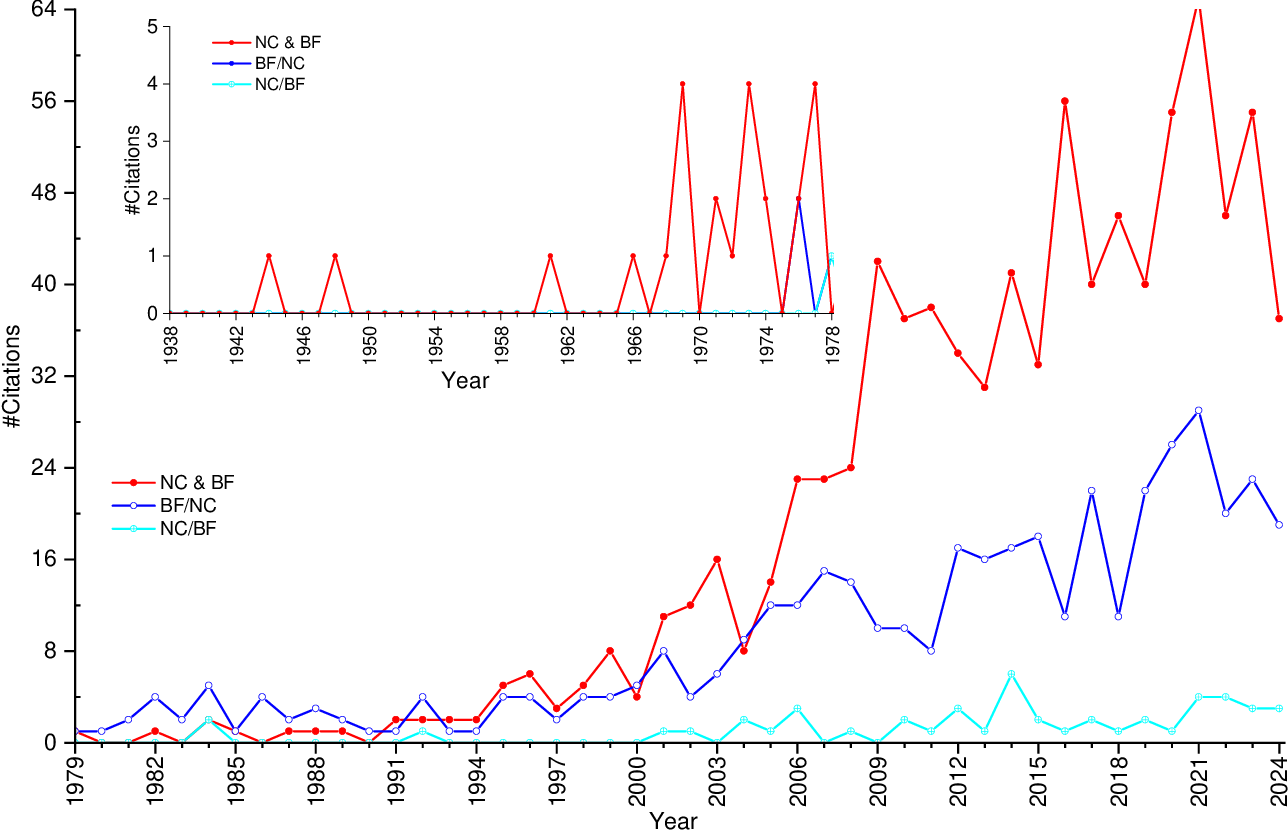}}
\vspace*{-10pt}
\caption{Comparison of yearly citations received by Benford (1938) only (BF/NC), Newcomb (1881) only (NC/BF) and both (BF \& NC). Insert plot is for years 1938-1978.}
\end{figure}

\section{Discussion}

When BF appeared on the scene NC with a lone citation it had received ironically from an experimental psychologist (Boring, 1920) had long faded into oblivion.  
Apparently unaware of its existence BF does not cite NC\footnote{Both Newcomb (1881) and Benford (1938) do not have any bibliography.}. However, NC continued to be neglected for another long spell of thirty eight years up to year 1976, a period which had seen the appearance of forty six articles on the phenomenon. BF, having got a first mover advantage after its chance pick up by Goudsmith and Furry (1944), had by then accumulated a comparatively respectable count of eighteen citations. Interestingly while it had taken psychologists pretty long forty years to notice NC BF caught their eye early on within ten years of its appearance when the observation was shown to be invalid for numbers invented by humans (Hsu, 1948).

Pinkham (1961), the most important contribution after Goudsmith and Furry (1944), proved the scale invariance of the law and further showed BL to be the only scale-invariant distribution of the first digits. 
It is seen from Table 2 that Pinkham (1961) with 346 citations is the fourth most cited paper on BL. Furthermore, Feller (1966) cited BF following Pinkham (1961) to explain the phenomenon, giving it a one page coverage, in the second volume of his book on probability theory\footnote{
Feller acknowledges personal communication with Pinkham.}.
Fairthorne (1969), another important paper citing BF after Goudsmith \& Furry (1944), 
is the most cited article in the field
of documentation. 
Thus Goudsmith \& Furry (1944) kept BF visible during the first three decades of its appearance leading to its timely citations by important papers, -  
none of which 
though considered NC.   
\begin{table}[h]
\rotatebox{270}{
\begin{minipage}{8.5in}
 \tbl{Important papers in the developmental history of BL arranged in order their citations}
{\begin{tabular}{@{}ccccccccccc@{}} \toprule
\hphantom{00} & &  & Cites & cocited & cocited & Cites  & cocited  & cocited & cocited &\\
No. & Paper \hphantom{00} & Citations  & Newcomb & with  & with  &  Benford & with & with & with &\\ 
\hphantom{00} &  &  & &  Newcomb &  Newcomb only &  & Benford & Benford only & both & \\ \botrule
1 & Hill (1995) \hphantom{00} & 672 & Yes & 453 & 12 & Yes & 559 & 116 & 441 &\\\\
2 & Nigrini (1996b) \hphantom{00} & 425 & No & 292 & 9 & Yes & 348 & 86 & 282 & \\\\
3 & Raimi (1976) \hphantom{00} & 359 & Yes & 245 & 6 & Yes & 300 & 62 & 232 & \\\\
4 & Pinkham (1961) \hphantom{00} & 346 & No & 265 & 6 & Yes & 316 & 58 & 259 & \\\\
5 & Nigrini \& \hphantom{00} & 323 & No & 194 & 5 & Yes & 263 & 75 & 187 \\ & Mittermaier (1997) \\\\
6 & Hill (1998) \hphantom{00} & 279 & No & 181 & 10 & Yes & 224 & 52 & 171 & \\\\
7 & Nigrini (1999) \hphantom{00} & 243 & No & 148 & 6 & Yes & 186 & 46 & 141 & \\\\
8 & Diaconis (1977) \hphantom{00} & 157 & No & 102 & 2 & Yes & 125 & 26 & 99 & \\\\
9 & Raimi (1969a) \hphantom{00} & 115 & No & 82 & 3 & Yes & 106 & 26 & 81 &  \\\\
10&Drake \&  \hphantom{00} & 114 & No & 76 & 2 & Yes & 97 & 24 & 74 \\ & Nigrini (2000) \\\\
11 & Hill (1988) \hphantom{00} & 74 & No & 58 & 0 & Yes & 71 & 13 & 58 & \\\\
12 & Goudsmith \&  & 44 & No & 31 & 1 &  Yes & 42 & 12 & 30 \\ & Furry (1944) \\\\ \botrule
 \hphantom{00} & Citations to Newcomb (1881) & 925 & & & & & Citations to Benford (1938) & 1325 & \\
 \botrule
\end{tabular} \label{ta3}}
\end{minipage}
}
\end{table}

NC continued to be dormant until 1976 when Cohen (1976) and Raimi (1976) papers
gave it a new lease of citation life.  
However, Cohen (1976), on BOB database, has a meager citation count of 37.  
Thus, we may consider that Cohen (1976) 
turned out to be of little consequence for the resurgence of the  FD  phenomenon. In fact, during our analysis, it was found that 
GS does not include Cohen (1976) within the search results of articles citing NC, - an exclusion that may have effected its visibility 
and in turn also of the latter.

An examination of the references showed that Cohen (1976) wrongly spells Newcomb, i.e., written  as \textbf{Necomb}, thereby  leading to its exclusion from search engine results. 
Raimi (1976), in stark contrast of Cohen (1976), has accumulated 359 citations (see Table 2);  it is currently the third most cited paper on the FD phenomenon.
NC, for the first time in about a century of its appearance, had achieved a yearly citation count of two,  though both of them are cocitations with BF, - which itself received four citations. Thus, 
1976 is a landmark year in the history of the  FD phenomenon, not only for Raimi's contribution to the mathematical understanding of the law and rediscovery of NC, but also for making the latter  
now onwards a contemporary of BF as far as their citation journey is concerned. 

At the time of Raimi (1976), NC had a total of three citations, whereas BF had a corresponding count of twenty-two. The respective citations increased somewhat disproportionately to 24 and 78 up to 1995 just before the publication of Hill (1995a),  which turns out to be another important paper citing both NC and BF. That citation gap between the two increased from 19 in 1976 to 54 in 1995, up to Hill (1995a), makes it clear that even after Raimi (1976) NC failed to get the attention that  should have been expected.  
The continuous neglect of NC is highly unanticipated in view of its new found, enhanced, visibility after Raimi (1976). Contrarily, one would have expected that NC would attract citations, at least, at par with BF. Thus arises the pertinent question: after Raimi (1976) what has led to a disproportionate number of citations of these two papers? In the treatment of NC and BF papers the sole thing which appears to be different was that Raimi (1976) formalized the name BL for the FD phenomenon;  only after Benford thereby as we stress, this was unjustly overlooking Newcomb.

The affirmative  answer to the foregoing question follows directly from an examination of the references in the papers published only the year after Raimi (1976). 
Two papers, Diaconis (1977) and Ylvisaker (1977), amongst a total of five, gave Raimi (1976) and BF quick co-citations; but surprisingly these two papers overlooked, or chose not to cite NC which therefore was denied two well deserved citations. The intentional disregard of NC motivated us to further investigate whether this peculiar citation behavior is an isolated aberration or is part of a larger pattern. 

The name Benford's law for the phenomenon was first coined in Raimi (1969a), an article in a popular magazine, Scientific American, which was a follow up of Raimi (1969b) when attempting a mathematical explanation of phenomenon through Banach and other scale invariances. Both Raimi's 1969 papers cite only BF whereas none cites NC or any other paper referencing it and therefore the author's ignorance of Newcomb's contribution can be safely assumed. Years later Raimi (1976) gave an extensive overview of  the literature and of different explanations of the phenomenon. However, too ironically Raimi (1976) formalized the name BL only after Benford though acknowledging, citing,  NC in the very same paper.

The first time the eponym BL appeared in the title of a scientific article is Wlodarski (1971). That FD phenomenon has now become synonymous with only Benford's name becomes clear when one considers how often the eponym BL is used in the title of the relevant publications. Out of a total of 2371 items indexed on BOB 1107 
use the name BL, a meager of 90 publications use Newcomb-Benford law whereas the term Newcomb's law is altogether non-existent in scientific parlance. Though the name BL itself speaks of the lack of acknowledgment of Newcomb's contribution, in order to quantify whether its formalization in Raimi (1976) effected the accumulation of future citations of the two papers differently, we inspected the bibliography of all the papers published from Raimi (1976) to Hill (1995a). 
The yearly data of the papers co-citing Raimi (1976) with BF/NC, NC/BF and NC \& BF is plotted in Fig. 3. It can be observed that during this period Raimi (1976) received 44 citations out of which
17 co-cite BF/NC, 10 NC \& BF, whereas only two papers co-cite Raimi (1976) with NC/BF.  
Raimi (1976) cited BF \& NC; therefore the authors of these 44 citing papers had knowledge of NC, and yet 32 of these chose not cite it. Either the authors deliberately did not cite NC or the formalization of the name ''BL'' overshadowed or diluted the extent of Newcomb's contribution. In our opinion, the latter assertion is much more plausible given that the inclusion of one additional reference is hardly laborious, unless the authors are sloppy enough. Overall up to the year 2024 Raimi (1976) is cocited with NC/BF merely six times whereas with BF/NC it is 62 times.

Furthermore, from  Fig. 3,  it seen that after 1995 BF \& NC always exceeds both NC/BF and BF/NC; this is an indication that the two papers are being invariably co-cited with Raimi (1976). This happened after a citation by Hill (1995a) brought NC back in focus and therefore imply the rationale for considering papers up to Hill (1995a); further disambiguation of the effect of the Raimi (1976) on the citations of NC from that of Hill (1995a) appears to be difficult.

\begin{figure}
\hspace*{-25pt}
\subfigure{\label{}\includegraphics[width=0.6\linewidth, height=1.1\linewidth,  angle=90,]{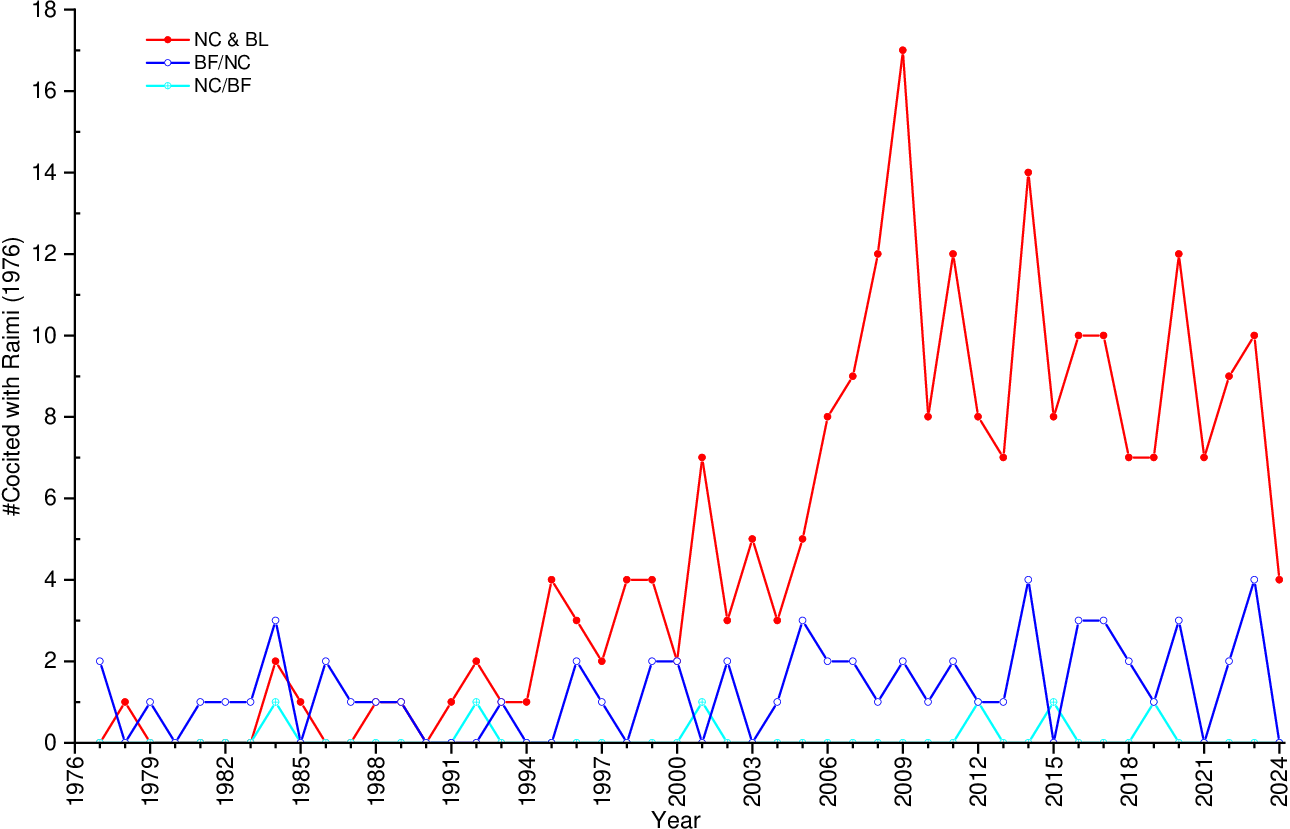}}
\vspace*{-10pt}
\caption{Comparison of yearly co-citations of Raimi (1976) with NC/BF, BF/NC and NC \& BF.}
\end{figure}

Diaconis (1977), determining the behavior of the first digits of arithmetic sequences from the theory of uniform distribution mod 1,    
despite referencing Raimi (1976), 
cites BF but not NC. The overlooking of NC continued further in Diaconis and Freedman (1979). In fact, Diaconis's knowledge of NC seems to have predated that of Raimi by many years. This can be concluded from an examination of the bibliography of the papers by Raimi (1976) and Cohen (1976) which shows Diaconis (1973) to be a reference common to both papers. Raimi (1976) also references Cohen (1976). It is obvious that the sole possible source of information of NC for Cohen (1976) is Diaconis (1973), since all other remaining references do not cite NC.
Diaconis (1977), amongst the most cited papers on BL, has 157 citations;  is co-cited with NC/BF only twice whereas it has 26 co-citations with BF/NC; with NC \& BF, it has 99 citations.  

The very sharp  increase in the number of publications in 1990s, as seen from Fig. 1, is indicative of an explosion of interest in BL. 
But it is dramatically obvious from Figures 1 and 2 that despite the rising popularity of the phenomenon the citation gap between the two papers is only widening further: the fame or notoriety for that matter, is not being equally shared by the two proponents of the law. When the two were woken up by the prince Hill (1995a), NC had 27 citations and BF had 83 citations, a gap of 56 which has increased to a larger value of 400, an increase of more than sevenfold, with former having 925 citations whereas latter having a respective count of 1325. The ever widening citation gap demonstrates the disproportionate appreciation of the two works,  even after being woken up by the same prince. 

T. P. Hill and M. Nigrini are the two most prolific authors on BL. The former made significant contributions to the theoretical understanding of the law, whereas the latter's ground breaking work ended up with  much skepticism on the practical usage of the BL. Hill, besides being a co-maintainer of BOB database, has authored (coauthored) 29 papers including a book on different mathematical aspects of the law. 
Hill's first paper on the phenomenon (Hill 1988) cited BF but not NC despite having Raimi (1976) in its bibliography. 
Hill (1988) has a total of 74 citations out of which it is never co-cited with NC/BF, 13 times with BF/NC, and 58 times with NC \& BF. Later on, in 1995, Hill co-cited NC and BF in all of his three papers which dealt with the mathematical foundations of the law. It is fair to stress here that Hill (1995a) proposed one of the most convincing explanations of BL through the use of statistical theory. 
Moreover, Hill (1995b) proved that base-invariance implies BL, another  
interesting  property of the law.  With 672 citations Hill (1995a) is currently the most cited paper of the author and also the most cited paper on BL, - after NC and BF papers, of course. In fact, this paper is being cited more often than NC; going by the current trends, it will soon overtake the latter to become the second most, after BF, cited paper in the field. Hill's second most cited paper ({\color{blue}{https://scholar.google.com/citations?user=LL0gUwkAAAAJ\&hl=en}}) Hill (1998) has 279 citations; it appeared in a popular science magazine, American Scientist. 
Surprisingly, it makes only a passing remark on the contribution of NC without its inclusion in the bibliography, - whereas BF received a proper citation. Hill (1998) is co-cited with NC \& BF 171 times, 52 times with BF/NC and has only 10 co-citations with NC/BF. Intriguingly, only a year latter, Hill (1999) in a paper in Chance magazine, cited both NC and BF.

M. Nigrini provided the first practical application of BL in the detection of financial misconduct, particularly the evasion of income taxes, receiving wide media coverage, thereby compelling the auditors and academicians to take note of this hitherto inconsequential empirical observation. There are a total of 34 publications including two popular books credited, authored or coauthored, to Nigrini on BOB.
An examination of the bibliographic information available for 24 of these publications reveals that only six cite NC, whereas BF is invariably cited. Nigrini's first publication on BL (Nigrini, 1992) is his Ph. D dissertation which cites both NC and BF. 
Further, Nigrini (1996a) again cited both papers.
Incomprehensibly, in the same year Nigrini (1996b) cited only BF but not NC. The latter paper, i.e. Nigrini (1996b), went on to become Nigrini's most acclaimed  paper ({\color{blue}{https://scholar.google.com/citations?user=glVvXlIAAAAJ\&hl=en}}) and is currently the second most cited paper on BL with 425 citations. This paper has 86 cocitations with BF/NC and only nine with NC/FB. Furthermore, being more frequently cited Nigrini (1996b), at the time of writing of Mir and Ausloos (2018), was the fourth most cited paper lagging behind Pinkham (1961) and Raimi (1976) which then were the second and third most cited papers on BL. 

The overlooking of NC continued further in Nigrini's second, third and fourth most cited papers (Nigrini \& Mittermaier, 1997; Nigrini, 1999, Drake \& Nigrini 2000) all of which again cite BF but none cites NC. Surprsingly only a year after Drake and Nigrini (2000), Nigrini (2001) cited both papers. The varying effect of the papers that have had the pivotal role in keeping the first digit phenomenon visible through various decades of its history is summarized in Table 2. The relevant citations data shown in Table 2 clearly indicates that  the lack of NC consideration in many of the most cited papers in the field, did  indeed adversely effect the future citations to NC.  

\begin{figure}
\hspace*{-25pt}
\subfigure{\label{}\includegraphics[width=0.6\linewidth, height=1.1\linewidth,  angle=90,]{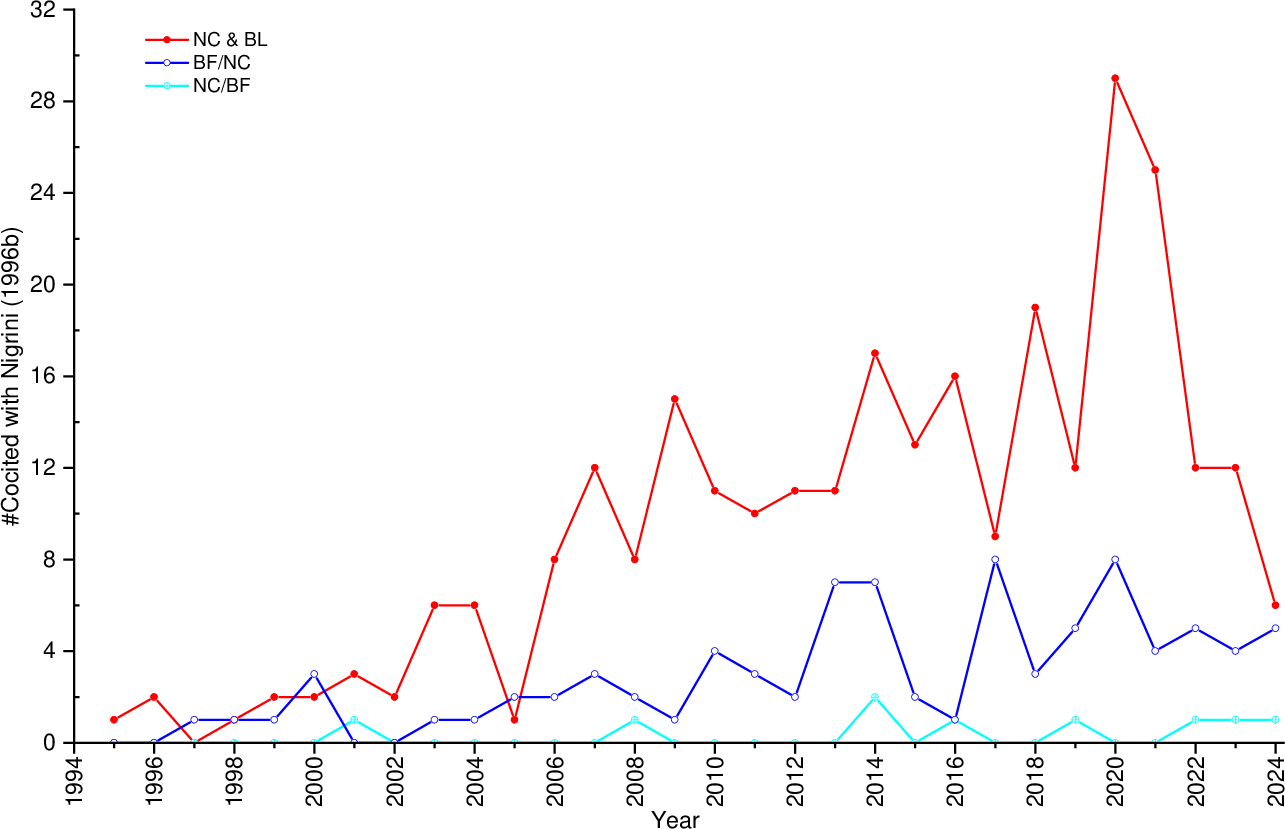}}
\vspace*{-10pt}
\caption{Comparison of yearly cocitations of Nigrini (1996b) with NC/BF,  BF/NC and NC \& BF.}
\end{figure}

\section{Conclusion}

In summary,
Newcomb (1881), after its initial obscurity, started its cocitation journey with Benford (1938) twice afresh, first after Raimi (1976) and then again after Hill (1995a); yet it did not receive the desired attention that would have been expected in view of the rising popularity of the FD phenomenon but rather continues to lag far behind 
Benford (1938) paper.

Newcomb (1881) is found to be placed in a journal between two articles of little scholarly worth and thus fate had it assigned a position 
that turned out to be strategically disadvantageous for its wider visibility. 
Then misfortune again struck Newcomb (1881) when Raimi (1976), its very own prince re-discoverer, overlooked it and formalized the name BL for the FD phenomenon only after  
Benford (1938), its citation contemporary, thereby unwittingly overshadowing 
Newcomb's contribution. 

Further, onwards
Newcomb (1881) stood little chance against Benford (1938) which, besides having the initial fortune of preceding Bethe et al. (1938), 
an important paper, had the  blessings of the prominent workers, with papers most cited in the field, who while taking great care in acknowledging (citing) the former were less considerate to and inconsistent in citing latter. 

We have therefore demonstrated how the number of citations of a paper can be very much altered as a function of its publication timing, its location in some scholarly (or not) journal, and how this number can depend on the citations (or not) by leading authors. This Newcomb (1881) vs. Benford (1938) "competition' or recognition can thus be considered as another type of Matthew effect.

In conclusion, let us observe that when one, a novice or a specialist of science, is told of BL immediate reaction is of shock and disbelief. The century long failure of mathematicians/statisticians to recognize a potential gem, longing for their attention in a purely mathematical journal, and in stark contrast the quick appreciation of the same, ironically by physicists, following its reappearance several decades later in a broad based journal adds further mystique to the 
saga of the enigma that is BL. Furthermore, over the years 
generations of stalwarts of science, trying their luck have walked down the treacherous path of unraveling the genesis of the FD phenomenon and yet only ``what'' it is about is agreed upon whereas the consensus on its why and how
remains elusive. 
Currently under intense scrutiny who knows what surprises BL has in store for its cult followers.




\end{document}